\documentclass[a4, 11pt]{article}
\usepackage{graphicx} % Required for inserting images
\usepackage{amsmath,amssymb,amsthm,amsfonts}
\usepackage{tikz}
\usepackage{pgfplots}
\usepackage{xcolor}
\usepackage{url}
\usepackage[round]{natbib}
\usepackage{float}
\usepackage{multirow}
\usepackage{csquotes}

\usepackage{tabularx}
\usepackage{booktabs}
\usepackage{subfigure}

\definecolor{blanchedalmond}{rgb}{1.0, 0.92, 0.8}
\definecolor{blanchedalmonddark}{rgb}{0.5, 0.92, 0.8}

\definecolor{DarkGreen}{rgb}{0.075,0.375,0.075}
\definecolor{DarkRed}{rgb}{0.5,0.1,0.1}
\definecolor{DarkBlue}{rgb}{0.1,0.1,0.5}
\definecolor{Gray}{rgb}{0.2,0.2,0.2}

\newcommand{\cA}{\mathcal{A}}

\newcommand{\cQ}{\mathcal{Q}}

\newcommand{\bigCI}{\mathrel{\text{\scalebox{1.07}{$\perp\mkern-10mu\perp$}}}}

\DeclareMathOperator*{\E}{\mathbb{E}}

\usetikzlibrary{shapes,decorations,arrows,calc,arrows.meta,fit,positioning}
\tikzset{
    -Latex,auto,node distance =1 cm and 1 cm,semithick,
    state/.style ={ellipse, draw, minimum width = 0.9 cm,minimum height = 0.9 cm,inner sep=0.02cm, fill=white},
    point/.style = {circle, draw, inner sep=0.04cm,fill,node contents={}},
    bidirected/.style={Latex-Latex,dashed},
    el/.style = {inner sep=2pt, align=left, sloped}
}

\usepackage[small]{caption}
\usepackage[pdftex]{hyperref}
\hypersetup{
    unicode=false,          % non-Latin characters in AcrobatÂ¿s bookmarks
    pdftoolbar=true,        % show Acrobat toolbar?
    pdfmenubar=true,        % show Acrobat menu?
    pdffitwindow=false,      % page fit to window when opened
    pdfnewwindow=true,      % links in new window
    colorlinks=true,       % false: boxed links; true: colored links
    linkcolor=DarkBlue,          % color of internal links
    citecolor=DarkGreen,        % color of links to bibliography
    filecolor=DarkRed,      % color of file links
    urlcolor=DarkBlue,          % color of external links
    pdftitle={},
    pdfauthor={},
    pdfkeywords={performative power, performative prediction, online search, digital markets, consumer protection, anti-trust}, % list of keywords
}

\usepackage[margin=1.1in]{geometry}

\usepackage{titlesec}
\titlespacing*{\paragraph}{0pt}{2ex plus 1ex minus .2ex}{1em}
\theoremstyle{plain}
\newtheorem{theorem}{Theorem}

\theoremstyle{definition}
\newtheorem{definition}{Definition}
\newtheorem{assumption}{Assumption}
\theoremstyle{remark}

\usepackage{authblk}
\stepcounter{footnote}
\addtocounter{footnote}{-1}
\author[1,2]{Celestine Mendler-Dünner}
\author[3]{Gabriele Carovano}
\author[1]{Moritz Hardt}
\affil[1]{Max-Planck Institute for Intelligent Systems, Tübingen and Tübingen AI Center}
\affil[2]{ELLIS Institute, Tübingen}
\affil[3]{Italian Competition Authority\footnote{Opinions expressed in the paper do not necessarily reflect the opinions of the AGCM.}} 
\date{}                 
\title{An engine not a camera: \\Measuring  performative power of online search}

\begin{document}

\maketitle

\vspace{-0.7cm}
\begin{abstract}
The power of digital platforms is at the center of major ongoing policy and regulatory efforts. To advance existing debates, we designed and executed an experiment to measure the performative power of online search providers. Instantiated in our setting, performative power quantifies the ability of a search engine to steer web traffic by rearranging results. To operationalize this definition we developed a browser extension that performs unassuming randomized experiments in the background. These randomized experiments emulate updates to the search algorithm and identify the causal effect of different content arrangements on clicks. Analyzing tens of thousands of clicks, we discuss what our robust quantitative findings say about the power of online search engines, using the Google Shopping antitrust investigation as a case study. More broadly, we envision our work to serve as a blueprint for how the recent definition of performative power can help integrate quantitative insights from online experiments with future investigations into the economic power of digital platforms.
\end{abstract}

\section{Introduction}

At the heart of one of Europe’s most prominent antitrust case is a seemingly mundane question: How much can a search engine redirect traffic through content positioning? In 2017, the European Commission alleged that Google favored its own comparison shopping service by steering clicks away from search results towards Google’s own product comparison service. %\footnote{European Commission, AT.39740, \textit{Google Search (Shopping)}, 27.06.2017.} 
The technical centerpiece of the case was an ad-hoc data analysis about the position and display biases of Google search results. Google appealed the European Commission's charges, pointing to, among other arguments, methodological errors.\footnote{Case C-48/22 P, Google and Alphabet v Commission (Google Shopping), ECLI:EU:C:2024:14.}

The case is emblematic of a broader problem. Although urgently needed, there is currently no accepted technical framework for answering basic questions about the economic power of digital platforms. Lawyers, economists, and policy makers agree that traditional antitrust tools struggle with multi-sided platforms~\citep{stigler19,cremer2019competition}. Against this backdrop, a recently developed concept from the machine learning literature, called performative power~\citep{hardt2022performative}, suggests a way to mitigate limitations of existing antitrust enforcement tools. Performative power measures how much a platform can causally influence platform users through its algorithmic actions. By directly relating power to a causal effect, it sidesteps the complexities underlying conventional market definitions and offers a promising framework to integrate data and experimental methods with digital market investigations. Although the definition of performative power enjoys appealing theoretical properties, a proof of its practical applicability was still missing.~\looseness-1

\begin{figure*}[t!]
\centering
\includegraphics[width=0.91\textwidth]{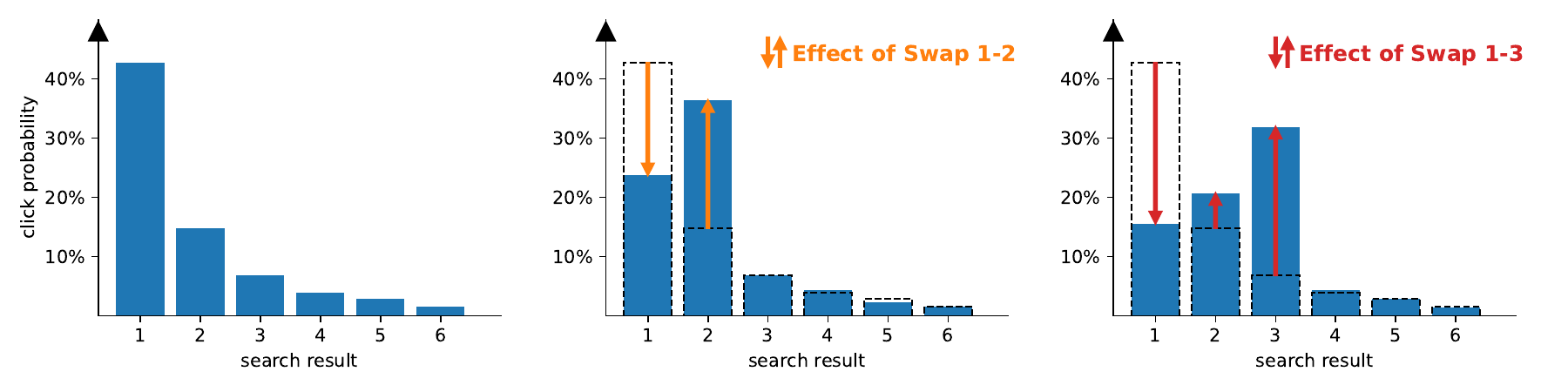}
\caption{The ability to influence web traffic through content arrangement. Blue bars show average click probability observed for generic search results in position 1 to 6 on Google search under different counterfactual arrangements; default arrangement (left),  swapping results 1 and 2 (middle),  swapping results 1 and 3 (right). We provide a detailed discussion in Section~\ref{sec:results} where we also explore arrangement changes beyond reranking.}
\label{fig:teaser}
\end{figure*}

\subsection{Our contributions}
In this work we present a first proof of concept showing how to use performative power as an investigative tool in practice.  
The instantiation of performative power we consider is motivated by the recent Google Shopping antitrust investigation ran by the European Commission against Alphabet Inc. It concerns the ability of a search engine to impact web traffic through decisions of how to arrange content.

Our core contribution is to design and implement an online experiment to establish a lower bound on performative power  by providing quantitative insights into the causal effect of algorithmic updates on clicks for the two most widely used search engines, Google Search and Bing. Our experiment is based on a browser extension, called Powermeter, that emulates  updates to the platform's algorithm by modifying how search results are displayed to users. We implement different counterfactual arrangements to inspect the effect of re-ranking and favored positioning (e.g., Ads or Shopping boxes) on clicks, both in isolation and jointly. The arrangement to which a user is exposed is chosen at random every time they perform a search. We discuss several technical steps we implemented to take care of the internal validity of our experimental design. 

Using Powermeter we collected data of about 57,000 search queries from more than 80 different subjects, over the period of 5 months. The queries for any given user follow the distribution of queries under their every-day use of online search. 
Figure~\ref{fig:teaser} provides a first glimpse into the observed effects of arrangement on clicks. In summary, we find that consistently down-ranking the first element by one position causes an average reduction in clicks of $42\%$ for the respective element on Google search. Down-ranking the same element by two positions yields a reduction of more than $50\%$. The causal effect is illustrated by the change in the hight of the bars. For Bing we find an even larger effect of ranking, although with less tight confidence intervals due to the small number of Bing queries performed by our participants. When combining down-ranking with the addition of Ads or Shopping boxes, the effect of arrangement is even more pronounced, showing a distortion in clicks for the first result of $66\%$ averaged across queries where such elements are naturally present on Google search.  Inspecting different subsets of queries we find that the effect of position is larger for queries with a high number of candidate search results. To the best of our knowledge, we are the first to offer independent quantitative experimental insights into display effects on Google search and Bing.

Finally, we outline how to formally relate our quantitative piece of evidence to questions about self-preferencing relevant in the context of the Google Shopping case. Taken together, we hope our empirical and theoretical results can serve as a first blueprint for what future antitrust investigations of digital platforms' market power based on performative power might look like.

\section{Preliminaries and related work}

The market power of digital platforms is the subject of a robust debate in policy, legal, and technical circles. See, for example, \citet{newman2014search,cremer2019competition, stigler19, comrep17uk, JRC122910}. Conventional antitrust enforcement tools have been put into question~\citep{syverson19mp} and new concepts of market power have been proposed to deal with the complexities of digital markets~\citep{oecdPowerConcept}. These account for the multi-sided nature of the markets as well as the role of behavioral weaknesses of consumers---albeit with limited success. We refer to a comprehensive literature survey about behavioral aspects in online market by the UK Competition and Market Authority~\citep{reviewUK}.

\paragraph{Performative power.} The concept of performative power offers an alternative notion of power inspired by recent developments in performative prediction~\citep{perdomo2020performative} within the computer science literature. 
We refer the reader to \citet{hardt2023performative} for a recent survey on the topic. A robust insight from performative prediction is that beyond learning patterns in data, the ability to \emph{steer} the data-generating distribution similarly factors into a predictive system's performance. 
Performative power~\citep{hardt2022performative} recognizes that the ability to steer depends on \emph{power}---in terms of reach and scale---of the platform making the predictions. Thus, the core idea behind performative power is to measure the degree to which predictions are performative to obtain an estimate of the power of a platform. Formally, 
performative power relates the ability of a platform to steer a population of participants, to the causal effect of algorithmic actions.~\looseness=-1

\begin{definition}[Performative power~\citep{hardt2022performative}] Given the  algorithmic action $a_0$ and a set of alternative conducts $\mathcal A$, a population $\cQ$ and an outcome variable $z$. Performative power is defined as 
\begin{equation}
\label{eq:defpp}
\mathrm{PP}:=\sup_{a \in \cA}\; \frac 1 {|\cQ|} \sum_{q \in \cQ} \E\|z_{a_0}(q)-z_a(q)\|_1\,,
\end{equation}
where $z_{a_0}(q)$ denotes the outcome for unit $q\in \cQ$ under $a_0$ and $z_a(q)$ denotes the counterfactual outcome, would the platform implement $a\in \cA$ instead. Expectations are taken with respect to the randomness in the potential outcome.
\end{definition}
Performative power is a measure of influence that predictive systems can have over their participants. It offers a family of definitions that can be instantiated flexibly in a given context. The specific meaning is determined by each instantiation. Performative power can be applied forward-looking to understand whether a platform has the ability to plausibly cause a specific change, as well as in retrospect to measure the effect of an observed conduct. In this work we use performative power to quantify the effect of an algorithmic update $a^*$ central to a recent antitrust investigation against Alphabet Inc.~ran by the European Commission.\footnote{European Commission, AT.39740, \textit{Google Search (Shopping),} 27.06.2017.\label{footnoteGS}} For this purpose, we instantiate $\cA$ with a set of conservative and implementable counterfactuals such as to provide a plausible lower bound on the effect of $a^*$.~\looseness=-1

\paragraph{The Google Shopping case.} In 2017 the European Commission imposed a fine of 2.42 billion EUR on Alphabet Inc.~for ``abusing its dominance as a search engine by favouring its comparison shopping service.''$^{\ref{footnoteGS}}$ The General Court dismissed Google's action against the decision and the Court of Justice of the European Union upheld the Court's ruling in 2024. It represents a landmark in EU competition law. The conduct under investigation concerned an update to the Google search algorithm. The update \textbf{a)} demoted rival comparison shopping services among general search results, often by multiple positions, and, at the same time, \textbf{b)} systematically gave prominent placement to Google's own comparison shopping service by triggering visually appealing boxes for shopping queries, reserved for Google's own products. The goal of this work is to provide quantitative insights into the effect of this conduct on web traffic by means of online experiments.~\looseness=-1

\paragraph{Display effects and behavioral biases.} Consumer choices on digital platforms are critically mediated by how platforms present content to users. Choice architecture designs~\citep{thaler2008nudge}, presentation bias~\citep{Yue10presentationbias}, position bias~\citep{joachmins05,guan07eye}, and trust bias~\citep{bing07googletrust} are known to play an important role. There is a rich literature in machine learning aiming to mechanistically understand such biases for debiasing click data~\citep{nick08,Chuklin2015ClickMF,guipeng22debias, jiwei23,agarwal2019general, zhen20LR, chen22ulr,yunan23bias}, building better ranking models and auctions~\citep{gagan06,1athey11positionauction}, and interpreting user feedback in recommender systems~\citep{joachmins05}, to name a few. Unfortunately behavioral aspects often resist a clean mathematical specification. By focusing on measuring a directly observable statistic, performative power circumvents the challenges of modeling behavioral biases for monitoring, auditing and measuring digital economies.

\paragraph{Measuring the effect of algorithmic updates.} Several prior works have been interested in measuring the effects of potential arrangement changes on consumption on online platforms. For example, \citet{ursu2018power} rely on public data collected under randomized result ordering to investigate the role of positioning on Expedia. \citet{narayana15ads} investigated position bias in search advertising using a regression discontinuity design. Also focusing on online advertising, \citet{agarwal11location} investigate position bias by experimentally randomizing bids to indirectly influence the ranking. Similarly in information retrieval researchers have studied active interventions in the form of order randomization~\citep{radlinski06random}, or relied on harvesting click data collected under multiple historical rankings~\citep{Fang19}. In our work we collect experimental data ourselves. We use a browser extension to emulate the algorithmic updates of interest \emph{without} requiring control over the platform's algorithm. 

Browser extensions have previously been used as a tool for automatically collecting data to audit systems. \citet{robertson-www-2018} audit Google search for polarization on politically-related searches. \citet{gleason23gatekeeper} collect data via an extension to directly investigate the effect of search result components on clicks. Also the ongoing National Internet Observatory~\citep{interObs} relies on a browser extension to collect web traffic data. While prior works focus on collecting observational data for monitoring systems, we use the extension to conduct online experiments.

\section{Performativity in online search}
\label{sec:setup}
We start by formalizing the causal question under investigation. We model an online search platform as a distribution over \emph{events}. An event is a triplet of a user query $Q$, content arrangement $A$ and click outcome $C$. A user query corresponds to a person visiting the search page and entering a search query in the search bar. The query is processed by the platform and results in an arrangement of content on the website. The mapping is typically defined by a proprietary pipeline involving a ranking algorithm that determines the order in which search results are ranked and displayed, including the positioning of components such as Ads or featured elements. 
Then, mediated by the arrangement, the user query leads to a click outcome $C$. The categorical random variable $C$ indexes the element clicked over by the user. It is a function of the user query and the arrangement.~\looseness-1

\subsection{The causal effect of arrangement}
Assume the platforms were to change their algorithm that determines the content arrangement. We seek to answer the following causal questions: \emph{How much does a change to the arrangement impact clicks of a content element on the search page?}

If clicks were solely determined by stable preferences, then we would see no effect. Performativity is the reason why we see an effect. Display baises, the limited ability to process large amounts of data, and trust in the platform can be a source for performativity. The more performative the arrangement is, the stronger the effect. We use $a_0$ to refer to the reference arrangement of results on Google search. For a given user query $q$ we define the potential outcome of a click event under the arrangement $a$ as $C_q(a)$. The variable~$C$ takes on categorical values, indexing the elements on the page. Let $\{c_1,...c_K\}$ denote the top $K$ general search results indexed in the order they appear under $a_0$.~\looseness-1
\begin{definition}[Performativity gap] Given a counterfactual arrangement $a'$, we define the \emph{performativity gap} at position~$i$ with respect to a population of queries $\cQ$ as \begin{equation*}\mathrm{\delta}^i(a')=\E\big[1\{C_q(a')=c_i\}\big]-\E\big[1\{C_q(a_0)=c_i\}\big]\,,
\label{eq:gap}
\end{equation*}
where expectations are taken over queries $q\in\mathcal Q$ and the randomness in the potential outcome.
\end{definition}
The performativity gap quantifies how much the click through rate of search item $c_i$ changed, in expectation across queries $\cQ$, had the platform deployed arrangement~$a'$ instead of arrangement $a_0$. 
The following result generalizes Theorem 8 in \citet{hardt2022performative}:

\begin{theorem}[Lowerbound on performative power]\label{thm:PP}
Let $\mathrm{PP}$ be the performative power of a search platform defined with respect to a set of arrangements $\cA$, a population of search queries $\mathcal Q$ performed on the platform, and the outcome variable $z_a(q) = \mathrm{1}\{C_q(a)=c_1\}$.
Then, performative power is lower bounded by the performativity gap as $\mathrm{PP}\geq\sup_{a\in\mathcal A}\delta^1(a)$. 
\end{theorem}

%It formalizes the idea that the average effect of an arrangement change on an individual query $q\in\cQ$ can be bounded by the average aggregate statistics across queries. 
Note that the instantiation of performative power in Theorem~\ref{thm:PP} to which we relate the performativity gap measures a platform's ability to steer \emph{outgoing} traffic from its online search website. We will discuss how to relate this notion to a broader discussion of the power of online search in vertically integrated markets in Section~\ref{sec:discussion}. 

\paragraph{Algorithmic distortion.} 
Often it can be useful to express the performativity gap relative to the base click rate. Thus, we define the \emph{algorithmic distortion factor} as the smallest factor $\beta>0$ such that 
\begin{equation}\mathrm{\delta}^i(a')\leq \beta \E\big[1\{C_q(a_0)=c_i\}\big].
\label{eq:dist}
\end{equation}
This quantity serves as a way to denote the fraction of clicks taken away from content item $c_i$ as a result of the update $a_0\rightarrow a'$. Also, as we will see, it helps to express performative power relative to a base click through rate which offers a more interpretable quantity for investigators.

\subsection{Estimating the performativity gap using an RCT} 
To estimate the performativity gap for the individual arrangements, we rely on a randomized controlled trial (RCT), the gold standard methodology to estimate causal effects~\citep{fisher1935, kohavi09web, imbens2015causal}. As we can not observe a search query simultaneously exposed to different arrangements, the idea of an RCT is to randomly select, for each query $q\in\cQ$, the arrangement they are exposed to. We write $Q_a\subset \mathcal Q$ for the subset of queries that are exposed to treatment $A=a$. We also refer to these subsets as  treatment groups. Comparing the click events across groups allows us to obtain an estimate of the performativity gap as 
\[{\mathrm{\bar \delta}}^i(a')=\mathrm{{CTR}}^i(a')-\mathrm{{CTR}}^i(a_0) \quad \text{with}\quad \mathrm{CTR}^i(a)=\frac 1 {|Q_a|} \sum_{q\in Q_a} \mathrm{1}\{C_q(a)=c_i\}.\]
For ${\mathrm{\bar \delta}}^i(a')$ to provide an unbiased estimate of $\delta^i(a')$, we rely on an application of the stable unit treatment value assumption (SUTVA)~\citep{rubin80sutva}, referred to as isolation assumption by~\citet{bottou2013counterfactual}: 
\begin{assumption}[Independence across queries]
User behavior in response to query $q$ is not affected by the treatment status of other queries, i.e., for all $q\in Q$ we have $C_q(A_q)\bigCI A_{q'}\;\forall q'\neq q $ where $A_q$ denotes the random variable assigning query $q$ to a treatment group.
\label{ass:indepQ}
\end{assumption}
This assumption justifies why we can interleave the measurement of different interventions in our experiment. It implies that the intervention performed on one query does not change individuals' browsing behavior in response to subsequent queries. For this to reasonably hold, individual interventions under should not impede user experience in a lasting manner. In the following we discuss steps we take in our experimental design towards justifying  Assumption~\ref{ass:indepQ}. 

More broadly, the key advantage of using an experimental approach to measure the performativity gap is that, while the mechanism mapping user queries to clicks can be arbitrarily complex, this complexity does not affect the experiment. Aspects such as users' preference for clicking links on the left side of the screen~\citep{dimitar16}, the effect of visually appealing elements~\citep{fubel2023rankings}, users' trust in the platform~\citep{bing07googletrust}, or the relevance gap between search results will naturally enter our measurement.

\section{Powermeter: Experiment in the wild}

We designed an online experiment to measure the performativity gap on two popular online search platforms operated by Google and Bing. The experiment is built around a Chrome browser extension that modifies the arrangement of search result pages and records user click statistics in a privacy-preserving fashion. The extension allows us to observe an organic set of user queries and click outcomes under different arrangements without having control over the platforms' algorithm.

\subsection{Browser extension}
Browser extensions can add functionalities to the web-browser and change how a website is displayed to the user. Powermeter makes use of these functionalities to emulate algorithmic updates by implementing different counterfactual arrangements on Google search and Bing search. We emphasize that Powermeter only hides or reorders, but never modifies or adds any content on the search page.

\paragraph{Technical details.} 
Once activated, the extension triggers the experiment whenever the user enters a search query on either Google search or Bing search. This can be identified by monitoring the url string of the current tab. Before search results are loaded the extension immediately hides the content of the website, inspects the html document, randomly assigns the user to one of the experimental groups and then implements the respective changes to the website before making the page visible. The implementation of the counterfactuals is done by identifying the relevant items to hide or swap by their html class names or ids. We also add custom tags and event listeners to the identified elements that we can fall back on at a later stage. 
The entire setup of the experiment usually takes around 40 milliseconds. This delay is far below what was found to be noticeable to users~\citep{brutlag2008user,arapakis14latency}. Hiding the html body of the website with the first possible Chrome event is crucial to avoid glitches in case of bad internet connection and make sure the control arrangement is not revealed to the participant. To ensure internal validity of our experiment, we also have to ensure a participant is never reassigned to a new experimental group when reloading a page, navigating between tabs or repeatedly entering the same search query. This is done by storing a hash of user ID and search query together with the assigned experimental group in the browser cache.

\paragraph{Backend and data collection.}
Every participant is assigned a unique random number that serves as anonymous user ID upon installation of the extension. This user ID persists throughout the experiment. Every time a click event on an element on the search page is registered, the click data is aggregated into a json object and sent to a database server hosted locally at out institution via a post request using the encrypted https protocol.
This concerns information about the index of the clicked search result, the click element type, the page index, and the experimental group. In addition, statistics about the website such as the number of search results, the presence of ads and boxes, the number of candidate results, and the position of specialized search results are extracted from the website are recorded.
The database server is built using the Microsoft~.Net core framework and deployed within a docker container. 
The database access is rate limited and the Get endpoint of the database is key protected. We use a SQlite database that is mapped to persistent memory.

\paragraph{Privacy considerations.} The information that is stored with every click does not contain any personally identifiable information. While we record the position of the clicked element on the search page, we never store search queries or any information about the websites visited by the user. This is an intentional choice to preserve user privacy, and to demonstrate that valuable insights can be gathered without privacy invasive data collection. The experiment went through an internal approval procedure and the privacy policy can be found on our website.\footnote{The project website can be found at: \href{https://powermeter.is.tue.mpg.de/}{https://powermeter.is.tue.mpg.de/}}

\begin{figure*}[t!]
\centering
\includegraphics[width=0.95\textwidth]{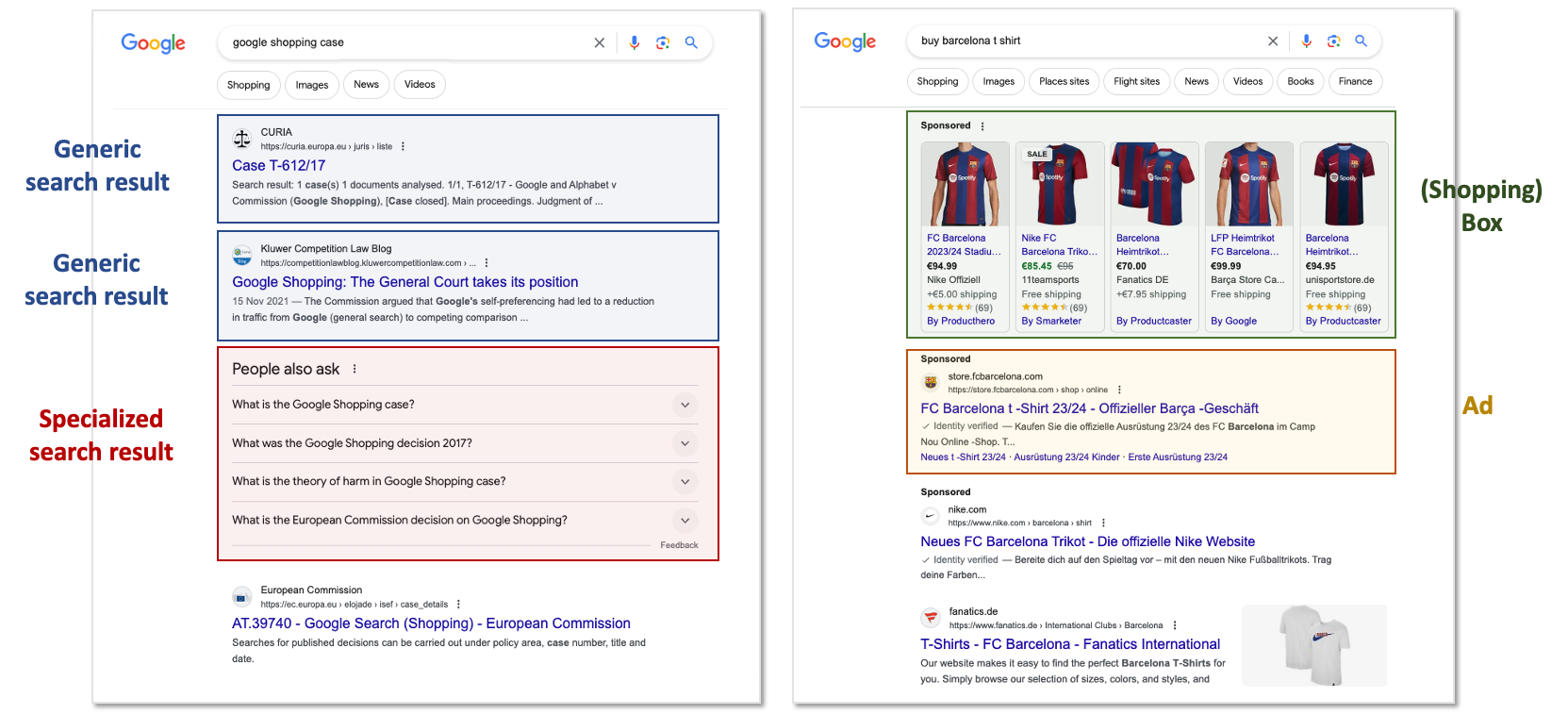}
\caption{Illustration of different elements on the Google search website.}
\label{fig:elements}
\end{figure*}

\begin{table*}[t!]
\centering
\caption{Counterfactual content arrangements implemented by Powermeter as part of the RCT.}
\small
\begin{tabular}{crl}\toprule
 \multicolumn{2}{r}{Arrangement}&Description\\ \hline
$a_0$ & control &Search results are displayed without any modification.\\
$a_1$ & swap 1-2 &The position of the first and the second generic search result are swapped.\\
$a_2$ & swap 1-3 &The position of the first and the third generic search result are swapped.\\
$a_3$ & swap 2-3 &The position of the second and the third generic search result are swapped.\\
$a_4$ & hide Ads/Box& Top Ads and Shopping boxes are hidden.\\
$a_5$ & hide + swap & Combines the latter modification ($a_4$) with swap 1-2 ($a_1$).\\
$a_6$ & hide Box & The shopping boxes are hidden.\\
\bottomrule
\end{tabular}
\label{tab:counterfactuals}
\end{table*}

\subsection{Experimental groups}
\label{sec:counterfactuals}
We implement six different counterfactual arrangements, summarized in Table~\ref{tab:counterfactuals}, each defining a treatment group. We refer to Figure~\ref{fig:elements} for the terminology used to refer to individual elements on the general search page. It equivalently applies to both, Google search and Bing. Arrangements $a_1-a_6$ are designed to emulate conservative variants of the conduct $a^*$ of interest, to inform a plausible lower bound on performative power. 
The first three arrangements $a_1$--$a_3$ concern the reordering of organic search results, leaving the other elements on the website untouched. The arrangements $a_4$ and $a_6$ perform modifications not directly concerning organic search results: Arrangement $a_6$  hides a specific element, called the Shopping box, appearing either in the right side panel or on top of the search results page. Arrangement $a_4$ hides the box together with all the Ads. Finally, Arrangement $a_5$ combines the latter change with a change in search result order. For Bing we only implemented the counterfactual $a_1$ to ensure statistical power despite data scarcity. 
A practical reason not to implement larger modifications is also users' sensitivity to the resulting deterioration of quality towards ensuring Assumption~\ref{ass:indepQ}. The Bing experiment of the European Commission's investigation had to be discontinued after one week for that exact reason.\footnote{See Case T-612/17, \textit{supra} 2, para 399} We made sure to avoid a similar failure point. Based on user feedback collected during an initial test round there was no indication that the modifications were even noticeable to users.

\subsection{Onboarding}

Participants were provided the link to the project website as an entry point.
The website contains information about the experiment, the purpose of the study, an onboarding video, as well as the privacy policy of the extension. The extension itself is distributed through the official Chrome webstore and there is a button directly navigating the user to the item in the store. We did not list the extension publicly to ensure participants are informed about the purpose of the study, and protect the integrity of our data. 
The installation follows the standard procedure of adding a browser extension to Chrome. The user has to give consent to access Google and Bing websites, as well as to use the storage API. 
The extension remains active until participants remove it from their web-browser, or until the experiment is stopped.  The study participants are trusted individuals of different age groups and backgrounds, recruited by reaching out personally or via email. We provide demographic statistics over our pool of participants in Figure~\ref{fig:Ustats} in the appendix.~\looseness=-1

\paragraph{Data preprocessing.} 
For each participant we ignored the clicks collected during the first four days after onboarding, as suggested by~\citet{keusch22awareness} in order to avoid potential confounding due to participation awareness.
We also removed clicks where the search elements could not be identified reliably for implementing the RCT to avoid selection bias towards the control group.~\looseness=-1

\section{Empirical results}
\label{sec:results}
Using our Powermeter browser extension we collected click data from $85$ participants over the course of 5 months, from September 2023 until January 2024. This resulted in $56,971$ click events, and a total of $45,625$ clicks after preprocessing. Out of the clicks $98.9\%$ were registered on Google, and $1.1\%$ on Bing. Figure~\ref{fig:stats} in the appendix visualizes some aggregate statistics over the clicks collected. We will consider several subsets of these events for which we measure the performativity gap and algorithmic distortion. In the following we discuss the main insights from the collected data. For all plots, we provide bootstrap confidence bounds over 200 resamples.

\subsection{Reordering search results}
\begin{figure*}[t!]
\centering
\subfigure{
\includegraphics[height=0.3\textwidth]{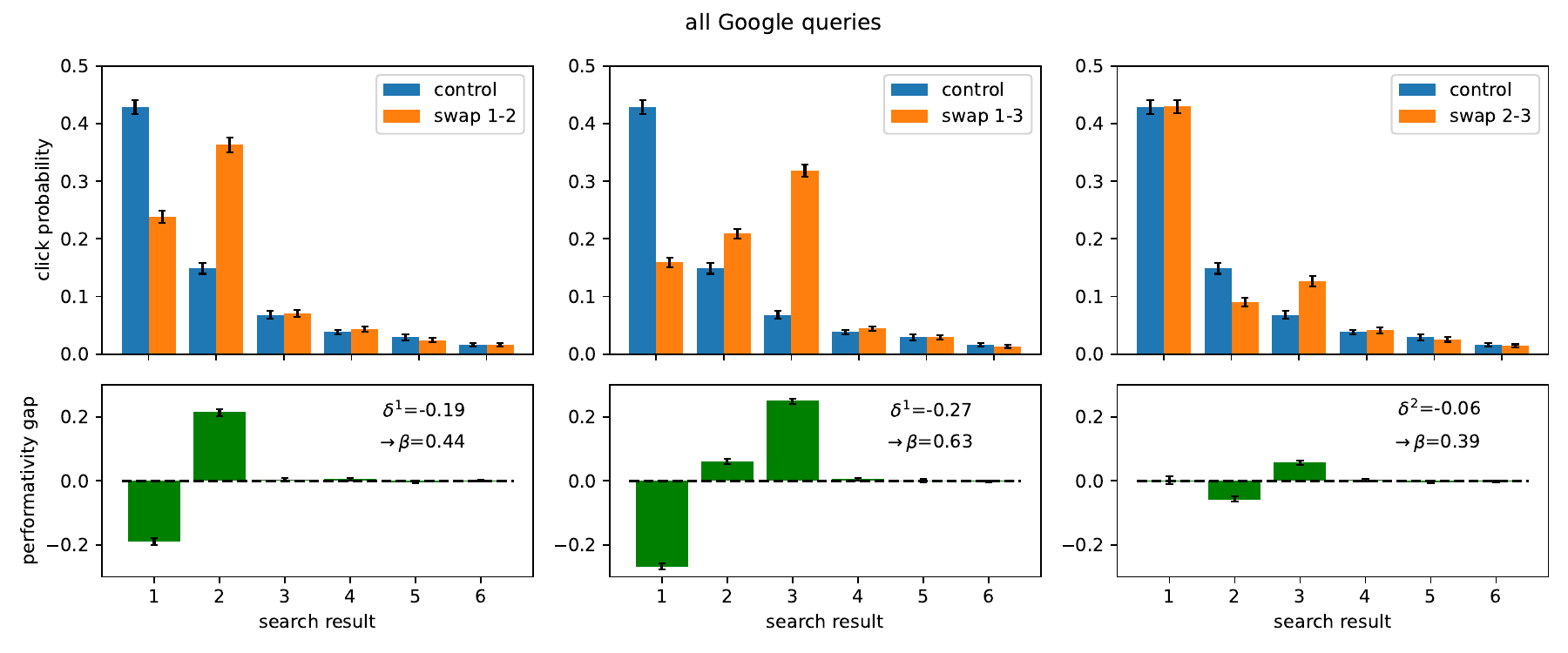}
}
\hfill
\subfigure{
\includegraphics[height=0.3\textwidth]{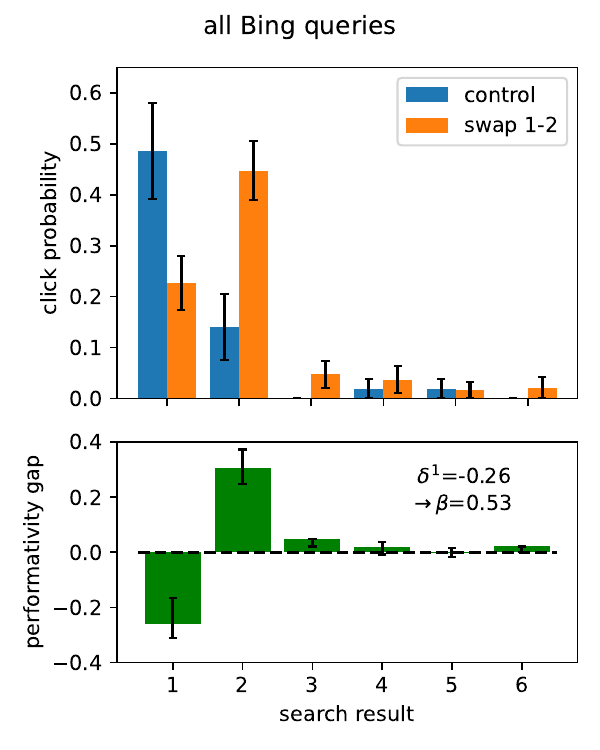}
}
\caption{Click through rate and performativity gap for general search results $c_1$ to $c_6$ under the counterfactual arrangements $a_1$, $a_2$, $a_3$ for Google and the counterfactual arrangement $a_1$ for Bing, compared to the control arrangement $a_0$ (in blue).}
\label{fig:ctr}
\end{figure*}

We first inspect the three counterfactual arrangements $a_1$, $a_2$, $a_3$ concerning reranking. Recall that $c_i$ indexes search results in the order in which they appears under $a_0$. In Figure~\ref{fig:ctr} we visualize the event probabilities $C=c_i$ for each search result $i=1,2,...,6$ under the control group (blue bars) and compare it to the respective probabilities under the three counterfactuals (orange bars). The figures on the left show the results across Google search queries. Here the counterfactuals correspond to swapping the position of the first two results (left), the first and the third (middle) and the second and the third (right). The right figure shows the results evaluated on Bing search queries when the first two results are swapped. The lower figure visualizes the corresponding performativity gap $\delta^i(a)$, corresponding to the change in clicks to item $i$ caused by the respective arrangement change.

We observe a consistently large effect of arrangement on clicks. Being down ranked by one position on Google search decreases average click through rate of $c_1$ from $43\%$ to $24\%$, resulting in $\delta^1(a_1)=-0.19$ and an algorithmic distortion of $\beta=0.44$. Being down-ranked by 2 positions results in $\delta^1(a_2)=-0.27$ and an average loss of more than $50\%$ of traffic. Note that a similar effect size has been reported in the case decision for the UK market for a two position shift~\cite[para 460]{GSdecision}, indicating that the estimate is robust across different user populations.
Finally, by down-ranking the second content element by one position we still observe a significant traffic reduction, corresponding to an algorithmic distortion of $\beta=0.39$. 

An interesting observation is also that for every counterfactual arrangement, the element shown first ends up getting most clicks on average. Implying that all the rankings are close to performatively stable~\citep{perdomo2020performative} with respect to the non-personalized reranking strategies considered. However, there are several indications of Google's ranking $a_0$ reflecting relevance of search results better then the other arrangements. Namely, $c_1$ gets more clicks when displayed first, compared to other results displayed in the same position ($c_2$ corresponds to first result under $a_2$ or and $c_3$ to the first result under $a_3$). A second indicator is that $c_2$ benefits from arrangement $a_3$; under the hypothesis that users consider search results in order this indicates that content item $c_3$ absorbs less clicks than $c_1$.

For Bing the position effect seems to be even more pronounced, although confidence intervals are significantly larger. We conjecture that this could be due to the average number of specialized search results and Ads present on the search page being larger on Bing. This results in a larger spacial separation of general search results and potentially larger display effects. Statistics about the type of elements present on the the search pages are reported in Figure~\ref{fig:stats} in the Appendix.

\subsection{Indirect effect of visually appealing elements}

\begin{figure*}[t!]
\centering
\includegraphics[width=0.325\textwidth]{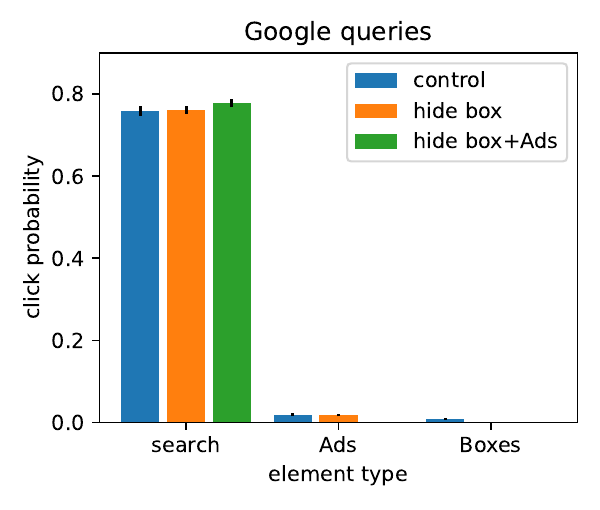}
\includegraphics[width=0.325\textwidth]{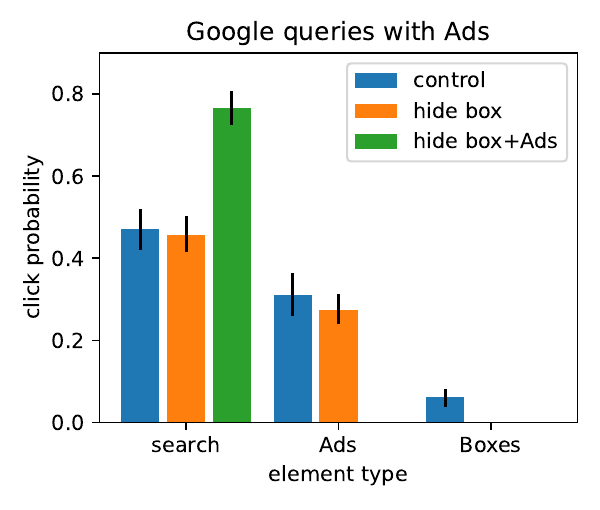}
\includegraphics[width=0.325\textwidth]{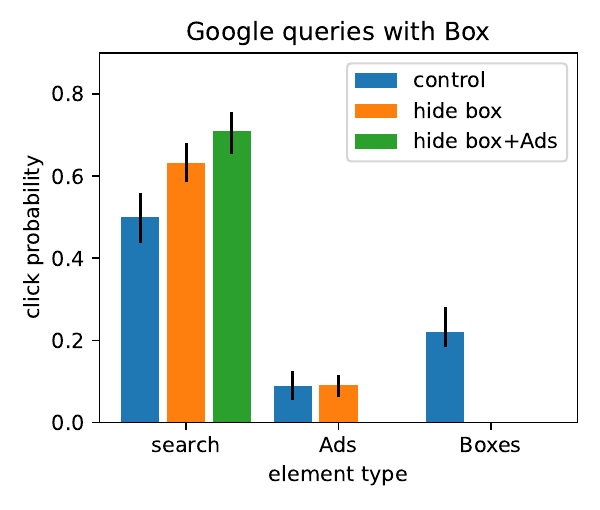}
\caption{Effect of arrangement on the click distribution across different element types (generic search results, Ads, boxes), visualized for three different subsets of Google queries. }
\label{fig:boxtype}
\end{figure*}

Next, we consider the counterfactuals $a_4$ and  $a_6$ that leave generic search results untouched and hide certain elements on the website. We first inspect the number of clicks these elements absorb if present on the page. We focus on Google search. In Figure~\ref{fig:boxtype} we compare the fraction of clicks going to generic search results, Ads, and Boxes for $a_0$, $a_4$, $a_6$. 
We plot the statistics across the aggregate queries (left panel), the subset of queries where Ads are present on the page under $a_0$ (middle panel) and the subset of queries where the box is present on the page under $a_0$ (right panel). We find the addition of boxes absorbs $22.4\%$ of the clicks on average across queries where it is present under $a_0$ and these clicks are mostly taken away form the generic search results. Similarly, Ads absorb close to $30\%$ of the clicks on average for queries where they are present. However, considering the overall number of clicks the effect is smaller since a large fraction of queries contains neither Ads nor Boxes.

\begin{figure*}[t!]
\centering
\begin{minipage}{0.6\textwidth}
\centering
\includegraphics[width=\textwidth]{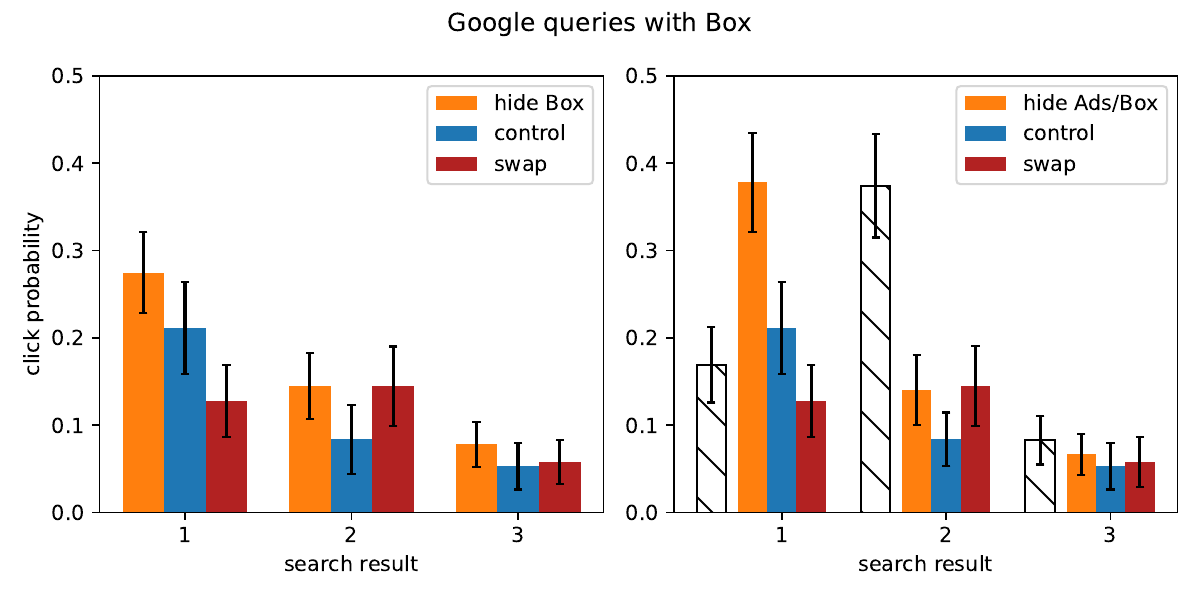}
\end{minipage}\hfill
\hspace{0.2cm}
\begin{minipage}{0.37\textwidth}
\small{
\begin{tabular}{| p{3.1cm} | c |}
 \hline
 {Conduct} & Distortion \\
  & of $c_1$\\
 \hline
 add Box&$\beta=0.23$\\
 add Ads/Box&$\beta=0.44$\\
 swap+add Box&$\beta=0.53$\\
 swap+add Ads/Box&$\beta=0.66$\\
 \hline
\end{tabular}}
\end{minipage}
\caption{Effect of hiding box and swapping elements on the click probability of generic search results. Statistics are evaluated for the subset of queries for which box is naturally present. The hashed bar shows the click probability under $a_5$ when top content is hidden and the first two elements are swapped. The right table reports the empirical measure of algorithmic distortion for different conducts, extracted form the results in the left figure.}
\label{fig:box}
\end{figure*}

\paragraph{Combined conduct.} We now consider the combined conduct of adding the box and down-ranking an element. We constrain our focus on queries where the Shopping box is present under $a_0$, either in the center column or in the right sidebar. These are $3.2\%$ of all the events. 
We show the corresponding click probabilities for the three first search results in Figure~\ref{fig:box}. In both figures the blue bars correspond to the control group $a_0$, and the red bars correspond to $a_1$. For these groups boxes are present on the page. In the left figure we investigate the effect of hiding boxes only and the orange bars correspond to arrangement $a_4$. In the right figure we investigate the effect of also hiding Ads, here the orange bars correspond to $a_6$.
We find that when adding Boxes, all three content items loose a significant fraction of clicks, whereas Ads mostly take away from $c_1$. In the right figure we additionally show $a_5$ using the hatched bars (i.e., down-ranking the first element by one position if box is hidden).
For $c_1$ the combined effect of adding Ads and Boxes on the click through rate is almost as large as the effect of down ranking the same item by one position. What we can see consistently is that combining the conduct of adding visually appealing elements on top of the page, and down-ranking a content item, has a combined effect that is larger than the effect of the two individual modifications alone. For element $c_1$ the measured distortion is reported in the table. The combined effect leads to a reduction of $25\%$ in clicks and an algorithmic distortion of $66\%$ when considering the effect of Ads/Boxes, and $53\%$ when only considering Boxes (comparing orange and red bar). We believe this to be the first time that quantitative insights into this combined conduct are made public.~\looseness=-1

\subsection{Factors that impact performative power} 

\begin{figure}[t!]
\centering
\includegraphics[height=0.225\columnwidth]{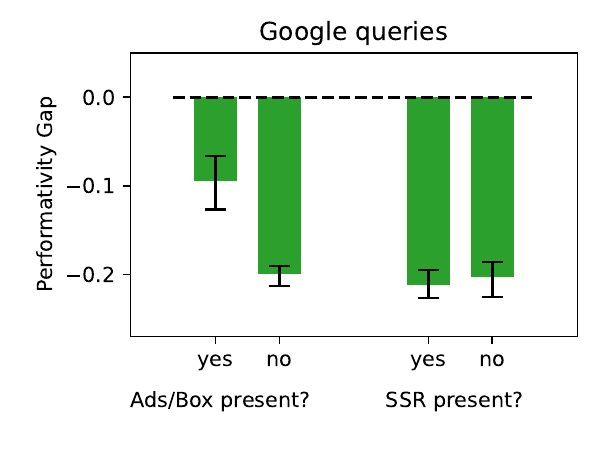}
\includegraphics[height=0.225\columnwidth]{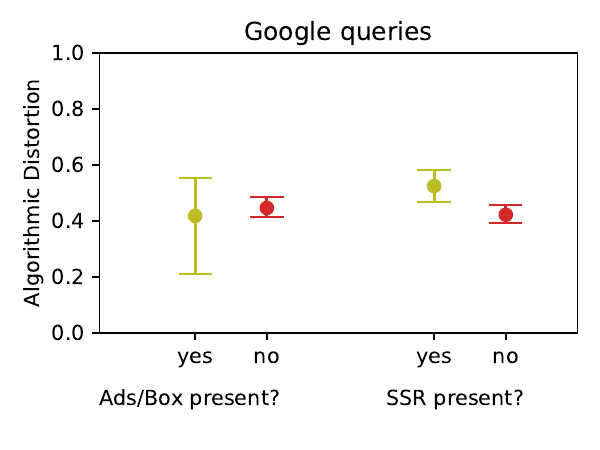}
\includegraphics[height=0.225\columnwidth]{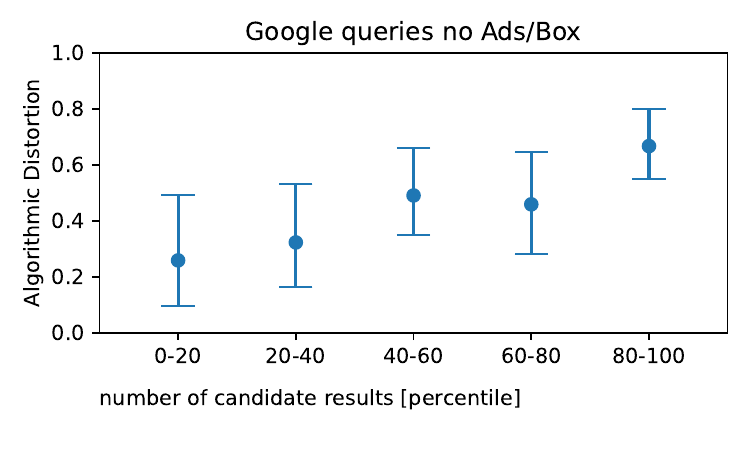}
\caption{Performativity gap and algorithmic distortion for content item $c_1$ under the counterfactual arrangement $a_1$ measure across different subsets of Google search queries. }
\label{fig:split}
\end{figure}

We perform additional investigations into what factors have a reinforcing effect on the performativity of ranking. To this end, we inspect different subsets of Google queries and measure the performativity gap as well as algorithmic distortion for $c_1$ under the counterfactual $a_1$. 
First, we split the data across two different axes depending on whether Ads or boxes are present, and whether Specialized Search results (SSR) are present in between the first two generic search results. The respective comparisons are visualized in the left and middle panel of Figure~\ref{fig:split}. We observe that the performativity gap in the presence of Ads and boxes is smaller, and about the same whether special search results are present in between search results.  However, if we plot distortion we get a different picture, since the base click probability of content item $c_1$ across different splits is different for the three cases. We find that while Ads and Boxes have little effect on distortion, the presence of a specialized search result in between the swapped results tends to  increase the effect of down-ranking $c_1$.

For the second investigation we group the queries by the number of candidate search results available on Google. This number was extracted from the website where it appeared as a string on top of the page in the form \texttt{`About 323'000'000 results (0.65 second)'}. The right panel of Figure~\ref{fig:split} shows algorithmic distortion for each percentile of the data. We see a clear trend that the performativity gap increases with the number of candidate results. We suspect this to be connected to the smaller relevance gap across results for queries with more potential results, leading to a higher influence of the arrangement on clicks. However, note that findings in this subsection are no causal claims.~\looseness=-1

\section{Discussion}
\label{sec:discussion}
We presented a flexible experimental design, based on a browser extension, to investigate the effect of search algorithms on user clicks. The browser extension performs interventions at the level of display to emulate algorithmic updates, without access to the platform algorithm. We implement different counterfactuals relevant for the Google Shopping antitrust investigation, and provide quantitative insights into the causal effect on clicks. Theorem~\ref{thm:PP} formally relates these quantitative findings to an instantiation of performative power, measuring the platform's ability to steer \emph{outgoing} traffic from its general search service.~\looseness-1

In a final step, we describe how our findings could fit with a broader anti-trust investigation potentially concerning effects spanning different markets. Take the Google Shopping case as an example. The case is concerned with the ability of Google search to distort \emph{incoming} traffic to a business operating in the market of comparison shopping services by changing where it appears on Google search relative to its competitors. Establishing the causal link between arrangements on search and their effect on incoming traffic to a third party website composes into two steps: a) establish Google's ability to steer outgoing traffic, b) quantify how much of the incoming traffic is mediated by Google search. The first step is a notion of power that experiments like ours operationalize, the second is a number that can readily be obtained from web traffic data. The overall performative power will be the product of the two factors. For putting this together, let's work though the following thought experiment:

\begin{quote}
\textit{Suppose, $80\%$ of the referrals to the competitor's website come from Google Search.\footnote{In 2022 up to $82\%$ of incoming traffic to Comparison Shopping Services in the European Economic Area was mediated by Google search, as reported in the Google Shopping case decision (Section 7.2.4.1, Table 24).} Further, assume that $70\%$ of the referrals from Google happen while the website is ranked among the top two generic search results. Our estimates suggest that distortion of traffic at the first position can be as large as $66\%$ for small arrangement changes. Assuming for the second position the effect is $20\%$ smaller, gives a conservative average effect size of $\beta=0.8\cdot 0.66$. Multiplying the effect size by the fraction of incoming clicks it concerns, we get $0.8\cdot 0.7\cdot \beta\approx 30\%$. This is the fraction of traffic to the site Google can redirect. }
\end{quote}

Turning this number into a conservative lower bound on performative power, it offers an interpretable measure of power for an investigator to judge whether in this case the algorithmic lever of self-preferencing should be a concern for competition in the down-stream market or not. We can use the same logic to compare search engines, and assess the effectiveness of remedies. 

More broadly, we hope our work can serve as a blue print for how performative power can be used to integrate experiments with future digital market investigations, and how tools from computer science, causality, and performative prediction can inform ongoing legal debates related to the power of digital platforms.
Our work is situated within a growing scholarship~\citep[c.f.,][]{metaxa21auditing,brandy21audit, dash2024antitrust} that takes advantage of the accessibility of digital markets for monitoring and regulating digital platforms. Beyond data, we demonstrate how experimental methods can offer an additional tool for power assessments.

From the perspective of computer science our work offers  measurements of performativity in the context of online search, contributing quantitative and empirical support to the study of performative prediction. As its name suggests a search engine is performative, it acts as an \emph{engine} steering consumption through its ranking algorithm, rather than a \emph{camera} merely picturing candidate results---we borrow this analogy from~\citet{mackenzie2008engine}.

\newpage
\section*{Acknowledgements}
We are particularly grateful to everyone who installed the extension and participated in our study; without you such a project would not have been feasible. Further, the authors would like to thank Jonathan Williams for the design of the project website, Alejandro Posada for producing the onboarding video and helping with the logo, Telintor Ntounis for assistance in setting up the server infrastructure, and Ana-Andreea Stoica, André Cruz and Jiduan Wu for feedback on the manuscript. We would also like to thank two anonymous reviewers at NeurIPS who provided valuable feedback related to the presentation and framing of our work. Celestine Mendler-Dünner acknowledges the financial support of the Hector Foundation.

\bibliographystyle{plainnat}
\bibliography{references}

\newpage
\clearpage

\appendix

\section{Limitations and Broader Impact}
\label{sec:limitations}

We develop a flexible methodology, to provide insights into the power of digital platforms. We hope our framework can support future digital market investigations, complement and address some of the limitations of current antitrust tools. This could help make firms accountable for steering behavior on digital platforms of various kinds, and potentially support cases of anti-trust, consumer protection and abuse of dominance in digital markets. Compared to traditional tools, our approach offers a more straights forward way to integrate experimental insights with regulatory questions and requires fewer assumptions on the market dynamics. Furthermore, by providing a quantitative measurement, the methodology also allows to compare platforms and assess the effectiveness of potential remedies. That being said, instantiating the definition in a meaningful way is still at the discretion of the investigator and requires substantial domain knowledge. Further, fitting the approach within existing legal frameworks is an open question, we hope to further make this concrete in future work.

On a technical side, our design ensures that for any given users, the observed clicks follow a natural distribution. However, our participants form a convenience sample of online search users. Depending on the target of the investigation this might not be sufficient to argue for external validity of the quantitative insights. We provide statistics about our user base to support such a judgement in Figure~\ref{fig:Ustats}. Further, we link these statistics with the observed clicks which can potentially help to adjust estimates using propensity score reweighting~\citet{stuart11reweightRCT}.  However, more rigorously arguing about the external validity of our results in specific contexts is left for future work. But we expect the qualitative take-aways of our work to generalize beyond our study, and hope they can inform future modeling choices and problem statements around performativity in online search. 

Lastly, we want to reiterate that our results for Bing should be taken with caution. While we designed our experiment to take most out of the available data, it is still a small sample of approximately $600$ search queries. Nevertheless, we did not want to withhold the results from the reader.

\section{Causal model} 

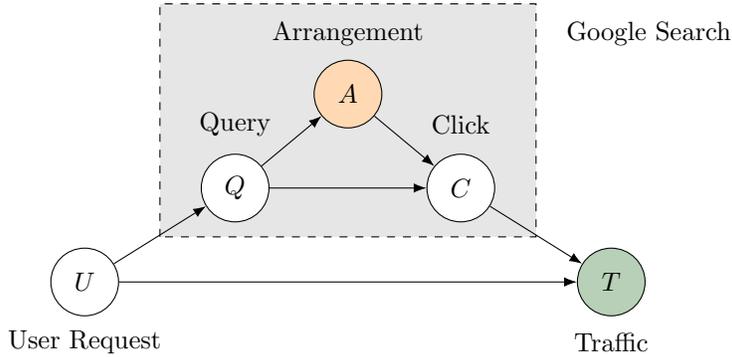
\begin{figure}[h!]
\centering
\small
\vspace{0.2cm}
\begin{tikzpicture}
    \draw[draw=black,dashed, fill=black!10!white] (0.5,-0.4) rectangle ++(5,3.1);
    \node[] at (7,2.3) {Google Search};
    \node[state,align=center] (R) at (-0.5,-1) {$U$};
    \node[] at (-0.5,-1.8) {User Request};
    \node[state,align=center] (C) at (4.5,0.25) {$C$};
    \node[] at (4.5,1.1) {Click};
    \node[state,align=center] (Q) at (1.5,0.25) {$Q$};
    \node[] at (1.5,1.1) {Query};
    \node[state,align=center, fill = orange!30!white] (A) at (3,1.5) {$A$};
    \node[] at (3,2.3) {Arrangement};
    \node[state,align=center, fill = DarkGreen!30!white] (T) at (6.5,-1) {$T$};
    \node[] at (6.5,-1.8) {Traffic};
    \path (R) edge (Q);
    \path (Q) edge (A);
    \path (A) edge (C);
    \path (Q) edge (C);
    \path (C) edge (T);
    \path (R) edge (T);
\end{tikzpicture}
\vspace{0.2cm}
\caption{Causal graph of online search users. A web request leads to the visit of a website, partially  mediated by Google search.} 
\label{fig:traffic}
\end{figure}

To support the arguments about composability of performative power in the discussion, consider the simplified causal diagram of how a user navigates to a website illustrated in Figure~\ref{fig:traffic}. 
The random variable $U$ denotes a user request and $T$ indicates which website is being visited in response. The user either navigates to the website via Google search (gray box), or they navigate to the website on some alternative way. This can be by entering the url directly, or using a different search service. 
Naturally, the arrangement $A$ only impacts the outcome $T$ if the user uses Google search, otherwise $A$ does not have any influence on the outcome. For every performed search query on Google, the user is exposed to an arrangement of content, and the arrangement mediates the click. Under appropriate independence assumptions the overall effect decomposes into two factors: the effect on outgoing traffic from Google search (gray box) and the overall fraction of relevant clicks being mediated by Google search. A formal treatment of this decomposition property under more general conditions is an interesting question for future work.~\looseness=-1

\newpage
\section{Proof of Theorem~\ref{thm:PP}}
\label{sec:proof}

We instantiate performative power with respect to a set of action $\mathcal A$, a population of search queries $\mathcal Q$, and the outcome variable $z_a(q) = \mathrm{1}[C_q(a)=c_i]$. Since the outcome is a scalar the $L_1$ norm reduces to an absolute value and we get 
\begin{equation}
\mathrm{PP}=\sup_{a \in \cA}\; \frac 1 {|\cQ|} \sum_{q \in \cQ} \E |z_{a_0}(q)-z_a(q)|
\end{equation}
Using the definition of the performativity gap and the definition of the $L_1$ norm the proof is a direct consequence of
\begin{align}
\mathrm{PP}&= \sup_{a \in \cA}\; \frac 1 {|\cQ|} \sum_{q \in \cQ} \E |\mathrm{1}[C_q(a_0)=c_i]-\mathrm{1}[C_q(a)=c_i]|\\ 
&\geq  \sup_{a \in \cA}\; \frac 1 {|\cQ|} \sum_{q \in \cQ}  \left|\mathrm{Pr}[C_q(a_0)=c_i]- \mathrm{Pr}[C_q(a)=c_i]\right|\\ 
&\geq  \sup_{a \in \cA}\;   \Big|\frac 1 {|\cQ|} \sum_{q \in \cQ} \mathrm{Pr}[C_q(a_0)=c_i]- \frac 1 {|\cQ|} \sum_{q \in \cQ} \mathrm{Pr}[C_q(a)=c_i]\Big|\\ 
&= \sup_{a\in\mathcal A} \delta^i(a)
\end{align}
where the performativity gap is defined with respect to the set of queries $\mathcal Q$. In words, Theorem~\ref{thm:PP} formalizes the idea that the average effect of an arrangement change on an individual query $q\in\cQ$ can be bounded by the average aggregate statistics across queries.

\newpage
\section{Additional details on the study}

\subsection{Aggregate click statistics}
\label{app:stats}
\begin{figure*}[h!]
\centering
\includegraphics[width=0.95\textwidth]{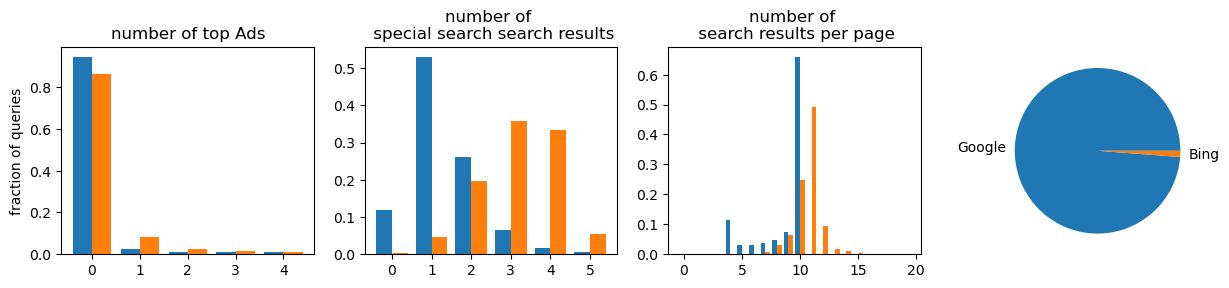}
\caption{Aggregate statistics over clicks and search result pages collected during our experiment. The blue bars show the statistics for Google and the orange bars show the statistics for Bing. Numbers are aggregated based on original search pages, before any modifications are performed. }
\label{fig:stats}
\end{figure*}

\begin{figure}[h!]
\centering
\includegraphics[width=0.8\textwidth]{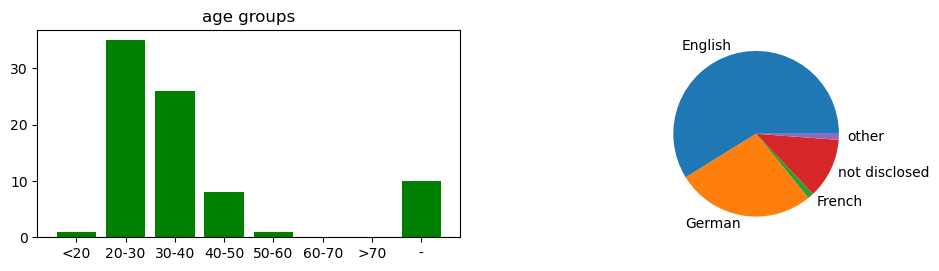}
\caption{Aggregate user statistics collected from the 85 participants with the onboarding form. Age distirbution (left) and language in which they consume online search (right). That's all the data we have about the demographics of our participants.}
\label{fig:Ustats}
\end{figure}

In Figure~\ref{fig:stats} we show aggregate statistics over the clicks collected. In Figure~\ref{fig:Ustats} we provide aggregate statistics over the user base. The latter information was collected through the onboarding form. 


\subsection{Project website}

\begin{figure}[h!]
    \centering
    \includegraphics[width = 0.48\textwidth]{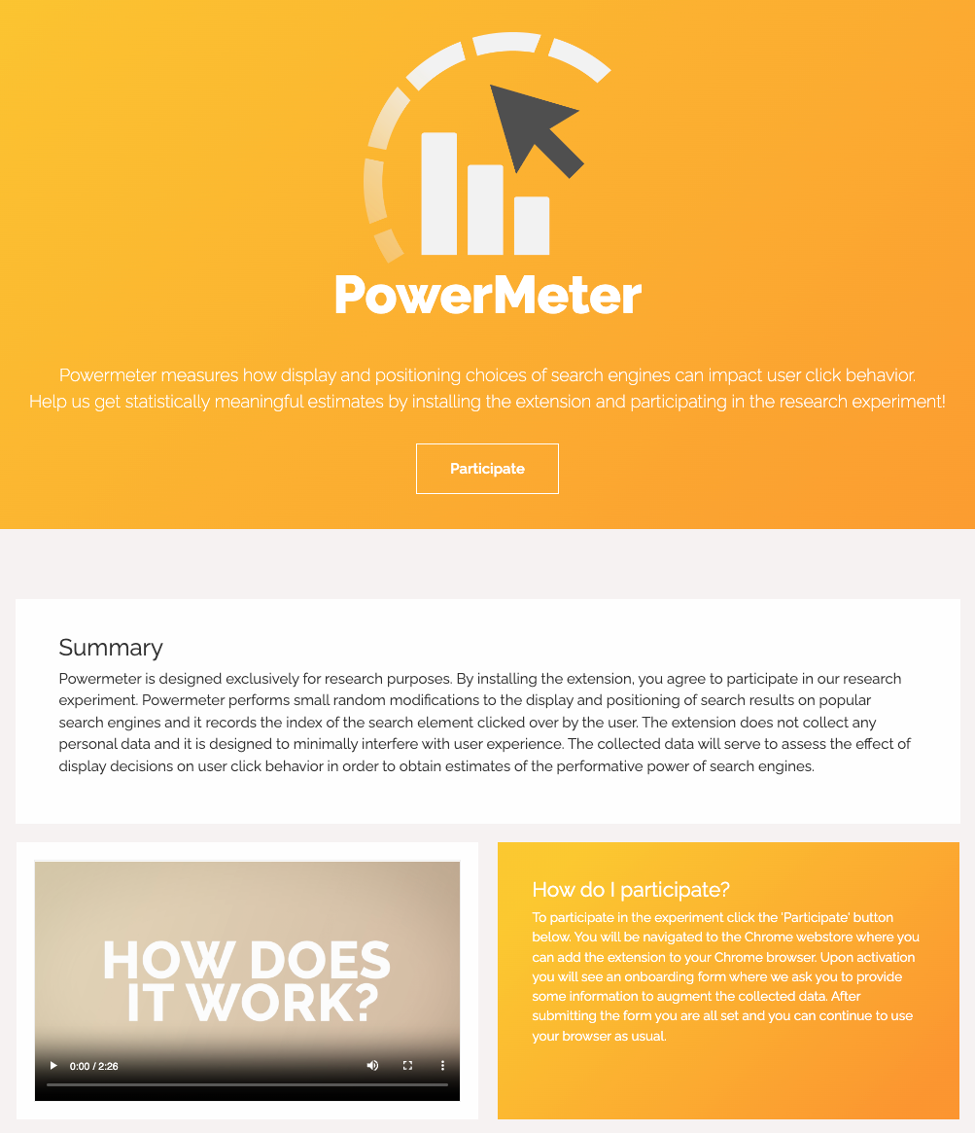}
    \includegraphics[width = 0.48\textwidth]{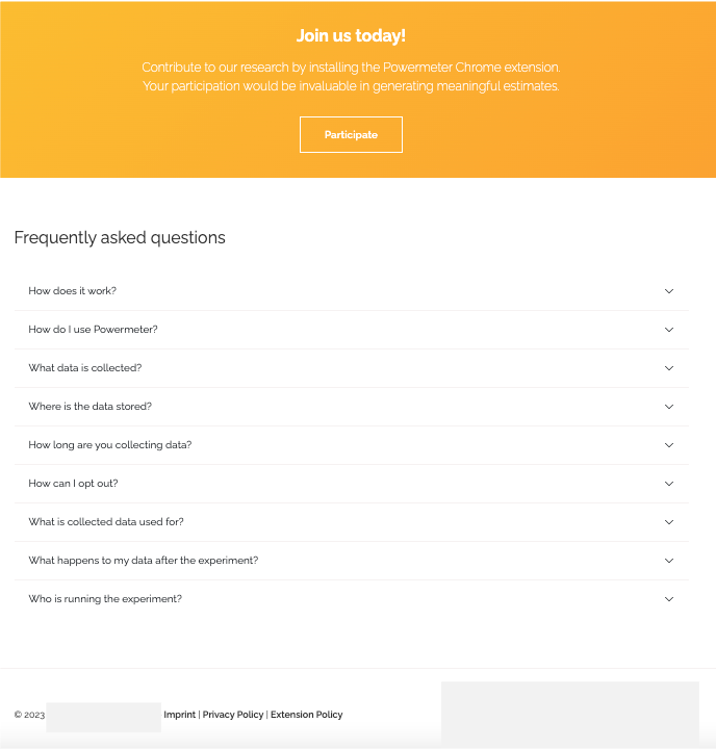}
    \caption{Project website. URL and institution names are removed for the sake of anonymity. }
    \label{fig:website}
\end{figure}

The project website provides the entry point to the experiment. Users where provided with details about the experiment, instructions for how to participate, and information about data usage. See Figure~\ref{fig:website} for a screenshot.

\subsection{Onboarding form}

\begin{figure}[h!]
    \centering
    \includegraphics[width = 0.48\textwidth]{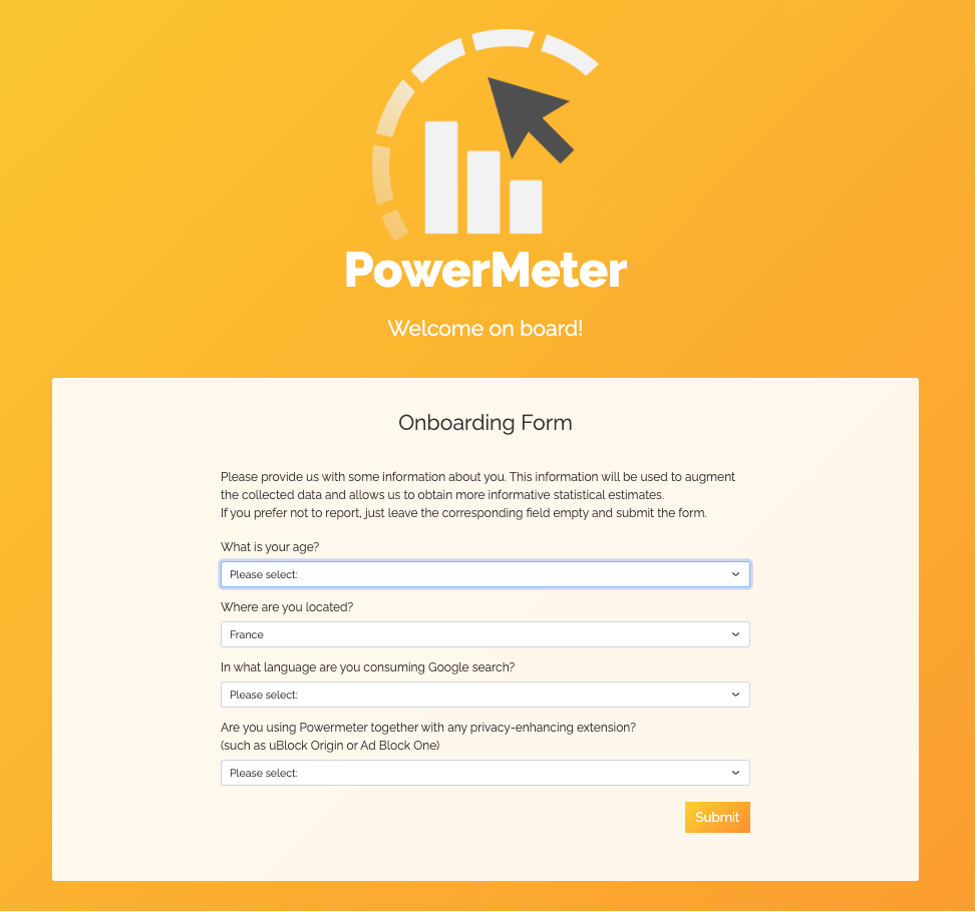}
    \caption{Onboarding form. }
    \label{fig:form}
\end{figure}

Upon installation of the extension the user is navigated to the onboarding form, as illustrated in Figure~\ref{fig:form}. Providing the information is not mandatory and answers are binned to only provide coarse grained information and no personally identifiable information. The main purpose of the information serves debugging different languages and website versions.

\end{document}